\documentclass[11pt]{article}

\usepackage[top=50mm, bottom=50mm, left=50mm, right=50mm]{geometry}
\usepackage{lineno}
\usepackage{amssymb}
\usepackage{amsmath}
\usepackage{amsthm}
\usepackage{epsfig}
\usepackage{graphicx}
\usepackage{graphics}
\usepackage{float}
\usepackage{subfigure}
\usepackage{multirow}
\usepackage{color}
\usepackage{lineno}
\usepackage{fullpage}
\usepackage[normalem]{ulem} 
\usepackage{makeidx}
\usepackage{xspace}
\usepackage{wrapfig}
\usepackage{algorithmicx}
\usepackage{algpseudocode}
\usepackage{forest}
\usepackage{url}
\makeindex

\newtheorem{theorem}{Theorem}

\newtheorem{lemma}{Lemma}

\newtheorem{procedure}{Procedure}

\newcommand{\EGZT}{Erd\H{o}s-Ginzburg-Ziv theorem}

\begin{document}

\title{Simple deterministic $\mathcal{O}(n \log n)$ algorithm finding a solution of Erdős-Ginzburg-Ziv theorem.}
 
\author{Seokhwan Choi, Hanpil Kang, Dongjae Lim}
\maketitle
\setcounter{page}{1}

\begin{abstract}
    \EGZT\ is a famous theorem in additive number theory, which states any sequence of $2n-1$ integers contains a subsequence of $n$ elements, with their sum being a multiple of $n$. In this article, we provide a deterministic algorithm finding a solution of \EGZT\ in $\mathcal{O}(n \log n)$ time. This is the first known deterministic $\mathcal{O}(n \log n)$ time algorithm finding a solution of \EGZT.
\end{abstract}

\section{\EGZT\ and its proof}

Let $EGZ(n)$ be a proposition which states for any sequence of $2n-1$ integers $A = \{{a}_{1}, {a}_{2}, \cdots, {a}_{2n-1}\}$, there exists a subsequence of size $n$, which we will call solution of theorem, such that their sum is divisible by $n$. \EGZT\ states $EGZ(n)$ holds for any positive integer $n$. \cite{1zbMATH03102822} We give an elementary proof of this theorem based on the original work of Erd\H{o}s, Ginzburg, and Ziv.

First, we will handle the case where $n$ is prime.

\begin{theorem}
$EGZ(p)$ holds for prime $p$.
\end{theorem}

Let $b_1, b_2, \cdots, b_{2p-1}$ be sorted sequence of $a_1, a_2, \cdots, a_{2p-1}$ according to their remainder divided by $p$. In other words, for some permutation $\pi$ of $\{1, 2, \cdots, 2p-1\}$, $a_{\pi(i)} = b_i$ ($1 \le i \le 2p-1$) and $(b_j \bmod p) \le (b_{j+1} \bmod p)$ ($1 \le j \le 2p-2$). Here, remainder of $x$ divided by $y$ is denoted as $x \bmod y$, and $0 \le (x \bmod y) < y$.

If index $i$ ($1 \le i \le p-1$) exists such that $b_{i} \equiv b_{i+p} \pmod{p}$, then $b_{i} \equiv b_{i+1} \equiv \cdots \equiv b_{i+p-1} \pmod{p}$. Therefore $\{b_i, b_{i+1}, \cdots, b_{i+p-1}\}$ is subset of $a$, with length $p$ and sum of elements in subsequence is divisible by $p$.

If such index does not exist, the difference $d_i = b_{i+p} - b_i$ is not multiple of $p$ for $1 \le i \le p-1$. We will consider submultisets of $A$ in the form of $\{c_1, \cdots, c_p\}$ with $c_j \in \{b_j, b_{j+p}\}$ for $1 \le j \le p-1$ and $c_p=b_p$. We can find a submultiset whose sum is multiple of $p$ among these subsets.

We will define a sequence of set $S_i$ ($1 \le i \le p$) as all possible remainders of sum of elements for the subsets, which follows the above constraints but $c_j=b_j$ for $j \ge i$. $S_i$ can be written as follows:

$$ S_i = \left\{\left(\sum_{j=1}^p c_j\right) \bmod p \ \middle|\ c_j \in \begin{array}{ll}
\{b_j, b_{j+p}\} & j < i \\
\{b_j\} & j \ge i
\end{array}
\right\}$$

We will find recursive formula for $S_i$. Here, we denote $X \oplus_p Y = \{(x+y) \bmod p \mid x \in X, y \in Y \} $.

\begin{lemma}
    For $2 \le i \le p$, $S_i =  S_{i-1} \oplus_p \{0, d_{i-1}\} $
\end{lemma}

\begin{align*}
S_i  & = \left\{\left(\sum_{j=1}^p c_j\right) \bmod p \ \middle|\ c_j \in \begin{array}{ll}
\{b_j, b_{j+p}\} & j < i \\
\{b_j\} & j \ge i
\end{array} 
\right\} \\ 
\ & = \left\{\left(\sum_{j=1}^p c_j\right) \bmod p \ \middle|\ c_j \in \begin{array}{ll}
\{b_j, b_{j+p}\} & j < i-1 \\
\{b_j\} & j = i-1 \\
\{b_j\} & j \ge i
\end{array}
\right\} \\ 
\ & \phantom{=} \cup \left\{\left(\sum_{j=1}^p c_j\right) \bmod p \ \middle|\ c_j \in \begin{array}{ll}
\{b_j, b_{j+p}\} & j < i-1 \\
\{b_{j+p}\} & j = i-1 \\
\{b_j\} & j \ge i
\end{array}
\right\} \\
\ & = S_{i-1} \cup \left\{\left(\left(\sum_{j=1}^p c_j\right) + b_{i+p-1}-b_{i-1}\right) \bmod p \ \middle|\ c_j \in \begin{array}{ll}
\{b_j, b_{j+p}\} & j < i-1 \\
\{b_j\} & j = i-1 \\
\{b_j\} & j \ge i
\end{array}
\right\} \\ 
\ & = S_{i-1} \cup (S_{i-1} \oplus_p \{b_{i+p-1}-b_{i-1}\}) = S_{i-1} \oplus_p \{0, d_{i-1}\} 
\end{align*}

\qed

\vspace{\baselineskip}

Now, we will prove that $\lvert S_{i} \rvert$ is strictly increasing until it becomes $p$.

\begin{lemma}
    For non-empty subset $S$ of $\{0, 1, \cdots, p-1\}$ and integer $a$ which is not multiple of $p$, $S \subsetneq S \oplus_p \{0, a\}$
\end{lemma}

$S=S \oplus_p \{0\}$, so it is trivial that $S \subset S +_p \{0, a\}$ and it is sufficient to prove $S \neq S \oplus_p \{0, a\}$.

If $S = S \oplus_p \{0, a\}$, $s \in S \rightarrow (s+a) \bmod p \in S+_p \{0, a\} \rightarrow (s+a) \bmod p \in S$.

$S$ is non-trivial subset of $\{0, 1, \cdots, p-1\}$, so there exists $x, y \in \{0, 1, \cdots, p-1\}$ such that $x \in S$ and $y \not\in S$. For positive $k \equiv (y-x) a^{-1} \pmod p$, $x \in S \rightarrow (x+a) \bmod p \in S \rightarrow (x + 2a) \mod p \in S \rightarrow \cdots \rightarrow  \left((x+ka) \bmod p\right)  = y  \in S$, which is contradiction to $y \not\in S$. So, $S \neq S +_p \{0, a\}$ and $S \subsetneq S+_p \{0, a\}$. \qed



\begin{lemma}

For $1 \le i \le p$, $|S_i| \ge i$.

\end{lemma}

We will use mathematical induction on $i$.

For base case $i=1$, $S_i = \left\{\left(\sum_{j=1}^{p} b_j\right) \bmod p \right\}$, so $|S_1| = 1 \ge 1$.

For inductive case $i \ge 2$, if $\lvert S_{i-1} \rvert = p$, then $S_{i-1}=\{0, 1, \cdots, p-1\}$ and also $S_i=S_{i-1} \oplus_p \{0, d_{i-1}\}=\{0, 1, \cdots, p-1\}\oplus_p \{0, d_{i-1}\}=\{0, 1, \cdots, p-1\}$. Therefore $|S_i|=p \ge i$.

If $|S_{i-1}| \ne p$, $S_{i-1} \subsetneq S_{i}$ by Lemma 2 and therefore $|S_{i-1}| < |S_{i}|$. So $|S_i| \ge |S_{i-1}|+1 \ge (i-1)+1 = i$. \qed

\vspace{\baselineskip}

$|S_p| \ge p$ by Lemma 3, so $|S_p| = p$ and $S_p = \{0, 1, \cdots, p-1\}$. This implies $0 \in S_p$, and proves existence of subsequence of length $p$, whose sum is divisible by $p$. \qed

\vspace{\baselineskip}

Now, we will handle the case where $n=pq$ is composite.

\begin{theorem}
$EGZ(p) \wedge EGZ(q) \rightarrow EGZ(pq)$
\end{theorem}

We will find $2p-1$ disjoint subsets $B_1, \cdots, B_{2p-1}$ of $A$ with $|B_i| = q$ and sum of elements in $B_i$ is multiple of $q$. $A \setminus (\bigcup_{j=1}^{i-1} B_{j})$ is a subset of size $q(2p-(i-1))-1$. If $i \le 2p-1$, $q(2p-(i-1))-1 \ge 2q-1$, so $EGZ(q)$ can be used to find $B_i$.

We will construct sequence $s$ of length $2p-1$, where $s_i = \frac{\sum B_i}{q}$. By applying $EGZ(p)$, we can select distinct $p$ indices $k_1, \cdots, k_p$ where $p \mid \sum_{i=1}^{p} s_{k_i}$. $\bigcup_{i=1}^{p} B_{k_i}$ is subset with size $pq$ and $\sum \bigcup_{i=1}^{p} B_{k_i} = \sum_{i=1}^{p} B_{k_i} = q \sum_{i=1}^{p} s_{k_i}$, which is multiple of $pq$. \qed

\vspace{\baselineskip}

Now, we will prove \EGZT .

\begin{theorem}
$EGZ(n)$ holds for any positive integer $n$.
\end{theorem}

We will use strong mathematical induction on $n$. For base case, $EGZ(1)$ holds since we can select whole sequence as subsequence. Length of such subsequence is $1$ and every number is divisible by $1$.

For inductive case $n \ge 2$, if $n$ is prime, $EGZ(n)$ holds by Theorem 1. if $n = ab$ is composite ($2 \le a, b < n$) $EGZ(a)$ and $EGZ(b)$ holds by induction hypothesis, and $EGZ(ab)$ holds by Theorem 2. \qed

\section{Efficient algorithm finding solution of \EGZT.}

By implementing constructive proof directly, we can provide an algorithm finding solutions of \EGZT\ of a sequence. By improving the algorithm for the case where $n$ is prime, we can achieve an overall time complexity of $\mathcal{O}(n \log n)$. Improvement uses implicit management of set $S_{1}, S_{2}, \cdots, S_{n}$ and modified binary search. This improves the result of \cite{2DELLUNGO20092658} and \cite{doi:10.1137/1.9781611976496.5}. 

\subsection{$n$ is prime}

For prime $n = p$, we will manage a sequence of sets $T_i$ with $T_i \subset S_i$, $|T_i| = i$ ($1 \le i \le p$) and $T_{j-1} \subset T_j$ ($2 \le j \le p$). For $i=1$, $|S_1|=1$ so $T_1=S_1$. For $2 \le i \le p$, $|T_{i-1}| < |T_{i-1} \oplus_p \{0, d_{i-1}\}|$ by Lemma 2, so there always exists an integer $0 \le t_i < p$ such that $(t_i - d_{i-1}) \bmod p \in T_{i-1}$ and $t_i \not \in T_{i-1}$. Then, $T_{i-1} \subset T_{i-1} \cup \{t_i\}$, $|T_{i-1} \cup \{t_i\}|=|T_{i-1}|+1=i$ and $T_{i-1} \cup \{t_i\} \subset T_{i-1} \oplus_p \{0, d_{i-1}\} \subset S_{i-1} \oplus_p \{0, d_{i-1}\} = S_i$. So we can construct $T_i$ as $T_{i-1} \cup \{t_i\}$.

To find $t_i$ efficiently, we will use modified binary search. For two integers $l < h$ such that $(ld_{i-1} \bmod p) \in T_{i-1}$ and $(hd_{i-1} \bmod p) \not \in T_{i-1}$, there should be an integer $x \in [l, h)$ with $(xd_{i-1} \bmod p) \in T_{i-1}$ and $((x+1)d_{i-1} \bmod p) \not \in T_{i-1}$. This can be shown in a similar way with the proof of Lemma 2.

For an integer $m \in [l, h)$, if $(md_{i-1} \bmod p) \in T_{i-1}$, there exists such $x$ in $[m, h)$. If not, there exists such $x$ in $[l, m)$, contrarily. We can reduce the range repeatedly until a single integer $x$ remains. Then $(x+1)d_{i-1} \bmod p$ can be chosen as $t_i$.

\begin{procedure} \texttt{Find\_t} finds $t_i$ to construct $T_i$.

\begin{tabular}{ll}
Input. & $p$, $T_{i-1}$, $d_{i-1}, u, v$ with $u \in T_{i-1}$, $v \not\in T_{i-1}$. \\ 

Output. & $t_i$ with $(t_i-d_{i-1}) \bmod p \in T_{i-1}$ and $t_i \not \in T_{i-1}$  \\
\end{tabular}

\end{procedure}

\begin{algorithmic}[1]

\Function{Find\_t}{$p$, $T_{i-1}$, $d_{i-1}$, $u$, $v$}

\State $l \gets (u \times d_{i-1}^{-1} \bmod p), h \gets p + (v \times d_{i-1}^{-1} \bmod p)$

\While {$l+1 \neq h$}
    \State $m \gets \left\lfloor \frac{l+h}{2} \right\rfloor$
    \If {$(m d_{i-1} \bmod p) \in T_{i-1}$ }
        \State $l \gets m$
    \Else 
        \State $h \gets m$
    \EndIf
\EndWhile
\State \Return $h d_{i-1} \bmod p$
\EndFunction

\end{algorithmic}

Now we will prove \texttt{Find\_t} takes $\mathcal{O}(\log p)$ time. 

\begin{theorem}
    \texttt{Find\_t} is correct and has time complexity $\mathcal{O}(\log p)$.
\end{theorem}

After line 2, $0 \le l < p \le h < 2p, ld_{i-1} \bmod p = u$ and $hd_{i-1} \bmod p = v$. Here, $d_{i-1}^{-1}$ can be found in $\mathcal{O}(\log p)$ time using extended Euclidean algorihtm.

While loop from line 3 to 10 has loop invariant $l<h$ and $(ld_{i-1} \bmod p) \in T_{i-1}, (hd_{i-1} \bmod p) \not \in T_{i-1}$. Thus after exiting loop, $l+1 = h, (ld_{i-1} \bmod p) \in T_{i-1}, (hd_{i-1} \bmod p) \not \in T_{i-1}$. $ld_{i-1} = hd_{i-1} - d_{i-1}$, so $hd_{i-1} \bmod p$ meets conditions for $t_i$.

After one step of while loop, $h-l$ becomes $\left\lceil\frac{h-l}{2}\right\rceil$ or $\left\lfloor\frac{h-l}{2}\right\rfloor$. After $k = \left\lceil \log_2(h-l) \right\rceil$ steps of while loop, $h-l$ becomes less than or equal to $\left\lceil\frac{h-l}{2^k}\right\rceil = 1$, which means loop terminates after at most $k$ steps. By managing $T_{i-1}$ with array of length $p$, single set operation can be run on $O(1)$ time, so each step of while loop takes $\mathcal{O}(1)$ time.  $k \le 1 + \log_2(h-l) \le 1 + \log_2(2p) \in \mathcal{O}(\log p)$, so total procedure takes $\mathcal{O}(\log p)$ time. \qed

\vspace{\baselineskip}

If $0 \not \in T_i$, we construct $T_{i+1}$ using $(b_1 + \cdots + b_p) \bmod p \in T_0 \subset T_i$ and $0 \not \in T_i$. If $0 \in T_i$, we can find a submultiset whose sum is divisible by $p$.

There is a submultiset $C_j$ of $A$ whose remainder of sum divided by $p$ is $(t_i - d_{i-1}) \bmod p$. Also, by definition of $T_{i-1}$ and $S_{i-1}$, subset with $b_{i-1} \in C_j$ can be found.

Remainder of sum of elements in $C_i = C_j \setminus \{b_{i-1}\} \cup \{b_{i+p-1}\}$ divided by $p$ is $(t_i-d_{i-1} + b_{i+p-1} - b_{i-1}) \mod p = t$. This subset can be recorded efficiently as tree structure, using the fact that only single element of a multiset is changed. 

\begin{center}

\begin{forest}
  [
    {$i = 1$, $t_i = 1$, $C_{i} = \{0, 1, 6, 2, 7\}$}
    [
        {$i = 2$, $t_i = 4$, $C_{i} = \{3, 1, 6, 2, 7\}$}
        [
            {$i = 4$, $t_i = 2$, $C_{i} = \{3, 1, 4, 2, 7\}$}
        ]
    ]
    [
        {$i = 3$, $t_i = 3$, $C_{i} = \{0, 8, 6, 2, 7\}$}
        [
            {$i = 5$, $t_i = 0$, $C_{i} = \{0, 8, 6, 4, 7\}$}
        ]
    ]
  ]
\end{forest}

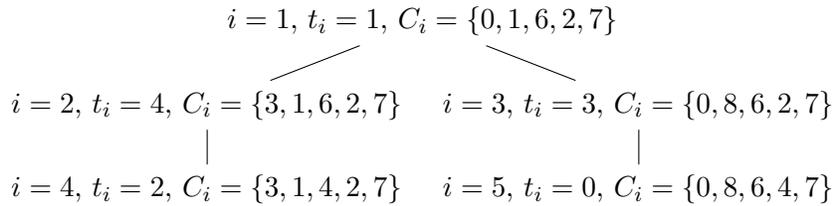
\textit{Figure 1. Tree structure storing submultiset $C$ when $p = 5$, $A = \{0, 1, 6, 2, 7, 3, 8, 4, 9\}$}

\end{center}

By implementing this strategy, we can calculate solutions of \EGZT\ for prime $p$.

\begin{procedure} \texttt{EGZ\_prime} calculating solutions of \EGZT\ when $n=p$ is prime.

Input. prime $p$ and array $a$ of $2p-1$ integers.

Output. $0$ or $1$ array $L$ of length $2p-1$, in which $\sum_{i=1}^{2p-1} {L}_{i} = p$ and $p \mid \sum_{i=1}^{2p-1} {a}_{i}{L}_{i}$.

\end{procedure}

\begin{algorithmic}[1]

\Function{EGZ\_prime}{$p$, $a$}

    \State $k \gets \Call{SORT}{[1, 2, \cdots, 2p-1], \texttt{key} = x \rightarrow a_x \bmod p}$
    
    \State $L \gets 0$-initialized array, indexed $1 \cdots {2p-1}$.
    
    \For {$i \gets 1 \cdots p$}
        \If {$a_{k_{i}} \bmod p = a_{k_{i+p-1}} \bmod p$}
            \State $L_{k_{i}}, L_{k_{i+1}}, \cdots,  L_{k_{i+p-1}} \gets 1$
            \State \Return $L$
        \EndIf
    \EndFor
    
    \State $s \gets (a_{k_1} + a_{k_2}+ \cdots + a_{k_p}) \bmod p$
    \State $T \gets$ \texttt{false}-initialized array, indexed $0 \cdots {p-1}$.
    \State $P \gets$ \texttt{NIL}-initialized array, indexed $0 \cdots {p-1}$.
    
    \State $T_s \gets \texttt{true}$

    \While {not $T_0$, \textbf{for} $i \gets 2 \cdots p$}
        \State $t_i \gets$ \Call{Find\_t}{$p, T, (a_{k_{p+i-1}}-a_{k_{i-1}}) \bmod p, s, 0$}
        \State $T_{t_i} \gets \texttt{true}$
        \State $P_{t_i} \gets i$
    \EndWhile
    
    \State $c \gets 0$
    \State $L_{k_{1}}, L_{k_{2}}, \cdots,  L_{k_{p}} \gets 1$ 
    \While{$s \neq c$}
        \State ${i} \gets {P_c}$
        \State $L_{k_{i+p-1}} \gets 1$
        \State $L_{k_{i-1}} \gets 0$
        \State $c \gets (c - (a_{k_{i+p-1}}-a_{k_{i-1}})) \bmod p $
    \EndWhile
    
    \State \Return L

\EndFunction

\end{algorithmic}

\subsection{$n$ is composite}

For composite $n = pq$, we can implement proof of Theorem 2 directly.

\begin{procedure} \texttt{EGZ\_composite} calculating solutions of \EGZT\ when $n=pq$ is composite.

Input. $p, q \ge 2$ and array $a$ of $2pq-1$ integers.

Output. $0$ or $1$ array $L$ of length $2pq-1$, in which $\sum_{i=1}^{2pq-1} {L}_{i} = pq$ and $pq \mid \sum_{i=1}^{2pq-1} {A}_{i}{L}_{i}$

\end{procedure}

\begin{algorithmic}[1]

\Function{EGZ\_composite}{$p$, $q$, $a$}

    \State $S \gets [1, 2, \cdots, p-1]$
    \State $T \gets $\texttt{NIL}-initialized array of length $2q-1$
    
    \For {$i \gets 1 \cdots 2q-1$}
        \State $S \gets S + [ip, ip+1, \cdots, (i+1)p-1]$
        \State $ret \gets \Call{EGZ}{p, [a_s | s \in S]}$
        \State $T_i \gets [S_j | ret_j = 1], S \gets [S_j | ret_j = 0]$
    \EndFor
    \State $L \gets 0$-initialized array of length $2pq-1$.

    \State $ret \gets \Call{EGZ}{q, \left[\frac{\sum_{t \in T_i} a_t }{p} \middle| i = 1, 2, \cdots, 2q-1\right]}$
    \For{$i$ s. t. $ret_i = 1$}
        \For{$j \in T_i$}
            \State $L_j \gets 1$
        \EndFor
    \EndFor
    
    \State \Return $L$
\EndFunction

\end{algorithmic}

\begin{procedure} \texttt{EGZ} calculating solutions of \EGZT.

Input. integer $n$ and array $a$ of $2n-1$ integers.

Output. $0$ or $1$ array $L$ of length $2n-1$, in which $\sum_{i=1}^{2n-1} {L}_{i} = n$ and $n \mid \sum_{i=1}^{2n-1} {a}_{i}{L}_{i}$

\end{procedure}

\begin{algorithmic}[1]

\Function{EGZ}{$n$, $a$}
    \If{$n = 1$}
        \State \Return [1]
    \EndIf
    
    \For {$i=2 \cdots n-1$}
        \If {$i \mid n$}
            \State \Return \Call{EGZ\_composite}{$i, \frac{n}{i}, a$}
        \EndIf
    \EndFor
    
    \State \Return \Call{EGZ\_prime}{$n, a$}
\EndFunction

Procedure 2 and 3 directly implements constructive proof of \EGZT.

\end{algorithmic}

\subsection{Time Complexity}

\begin{theorem}

procedure \texttt{EGZ} has time complexity $\mathcal{O}(n \log n)$.

\end{theorem}

Let's denote time complexity of $\texttt{EGZ}$ function as $T(n)$.

\texttt{EGZ\_composite} has time complexity $T(pq) = (2q-1)T(p) + T(q) + O(pq)$ where $2 \le p \le q$. For some constant $C$, $T(pq) \le (2q-1)T(p) + T(q) + Cpq$. \texttt{EGZ\_prime} calls \texttt{find\_t} at most $p-1$ times, resulting total time complexity of $T(n) = \mathcal{O}(n \log n)$ time. For some constant $D$, $T(n) \le D n \log n - C n$.

We will prove $T(n) \le D n \log n - C n$ for $n \ge 2$. We will use strong mathematical induction on $n$.

Including base case $n = 2$, it is definition of $C$ and $D$ such that $T(n) \le D n \log n - C n$ for prime $n$. If $n = pq$ with $2 \le p \le q$ is composite, $T(pq) \le (2q-1)T(p) + T(q) + Cpq$.

$\frac{d}{dx} \frac{x \log x}{x-1} = \frac{x - \log x - 1}{(x-1)^2} \ge 0$ For $x \ge 2$, thus $\frac{x \log x}{x-1}$ is increasing, and $(q-1) p \log p - (p-1) q \log q \le 0$. So, $T(pq) \le D ( pq \log pq + (q-1) p \log p - (p-1) q \log q) - C (pq + q - p) \le D pq \log pq - C pq$. \qed 

\section{Implementation}

Implementation can be found in \url{https://github.com/ho94949/egz}.

\bibliographystyle{plain}

\bibliography{ref}

\end{document}